\documentclass[pra,onecolumn]{revtex4}%
\usepackage{epsfig}
\usepackage{amsmath}
\usepackage{graphicx}
\usepackage{bm}
\usepackage{amsfonts}
\usepackage{amssymb}

\begin{document}
\title{Ultralong distance coupling between asymmetric resonant microcavities}
\author{Fang-Jie Shu,$^{1}$ Chang-Ling Zou,$^{2,*}$ Wen-Cong Chen,$^{1}$ Fang-Wen Sun,$^{2}$}
\address{
$^1$ Department of Physics, Shangqiu Normal University, Shangqiu 476000, P. R. China \\
$^2$ Key Lab of Quantum Information, University of Science and Technology of
China, Hefei 230026, P. R. China}
\email{clzou321@ustc.edu.cn}
\date{\today}

\begin{abstract}
The ultralong distance coupling between two Asymmetric Resonant Microcavities
(ARCs) is studied. Different from traditional short distance tunneling
coupling between microcavities, the high efficient free space directional
emission and excitation allow ultralong distance energy transfer between
ARCs. In this paper, a novel unidirectional emission ARC, which shows
directionality $I_{40}=0.54$, is designed for materials of refractive
index $n=2.0$. Compared with regular whispering gallery microresonators,
the coupled unidirectional emission ARCs show modulations of resonance
frequency and linewidth even when the distance between cavities is
much longer than wavelength. The performances and properties of the
ultralong distance interaction between ARCs are analyzed and studied
by coupling mode theory in details. The ultralong distance interaction
between ARCs provides a new way to free-space based optical interconnects
between components in integrated photonic circuits.
\end{abstract}

\maketitle

\section{Introduction}

Atoms are bonded together to form molecules, which greatly enrich
our world. The same story is true for microcavities \cite{1}. Coupling
of two microdisks \cite{2,3,4,5} provide appealing characters, such
as high-Q unidirectional emission, low-threshold and single-frequency
laser. These bonded microcavities, known as the photonic molecule
\cite{6}\cite{key-1}, are formed relying on the bond interaction
which comes from the short distance evanescent coupling. Nevertheless,
the typical length of evanescent tail of optical modes reaching out
of the cavities is smaller than the wavelength, thus it is required
a subtle control on subwavelength interval between cavities for forming
of a designed photonic molecule. Therefore, the experiment difficulties
are increased and the applications are limited.

For example, Stock et al. \cite{7} presented recently an on-chip
integrated quantum optical system made up with two micropillars: one
micropillar is an electrically pumped microlaser which serves as the
integrated light source, the other micropillar is a microcavity containing
quantum dots which serves as quantum optical device. In this configuration,
the energy transfered from the light source to the device is crucial.
However, due to the fact that each pillar cavity is covered by an
electrode or a micro-aperture which much larger than themselves, the
gap between two cavities is naturally much longer than the wavelength,
thus the short distance interaction is very inefficient. As suggested
by the authors, the asymmetric resonant cavities (ARCs), instead of
regular circular shaped cavities, can be applied for efficient energy
transfer.

In this paper, we numerically demonstrated the ultralong distance
interaction between ARCs through radiation coupling. The key issue
of the long distance interaction is the high directionality emission
of ARC, which determines the light emission and collection efficiency
\cite{9,10,11}. First of all, we proposed a novel ARC for unidirectional
emission. Then, we carried out a strong far-field coupling system
composed by two ARCs placed as emission direction opposite to each
other. Next, numerical and theoretical studies are performed to investigate
the unidirectional emission character of individual cavity, coupling
induced resonance variation, and energy transferring between two cavities.
The ultralong coupling scheme provides a new energy saving way to
exchange energy on-chip.

\section{ARC with Unidirectional Emissions}

\begin{figure}
\label{fig1}\centerline{ \includegraphics[clip,width=0.9\columnwidth]{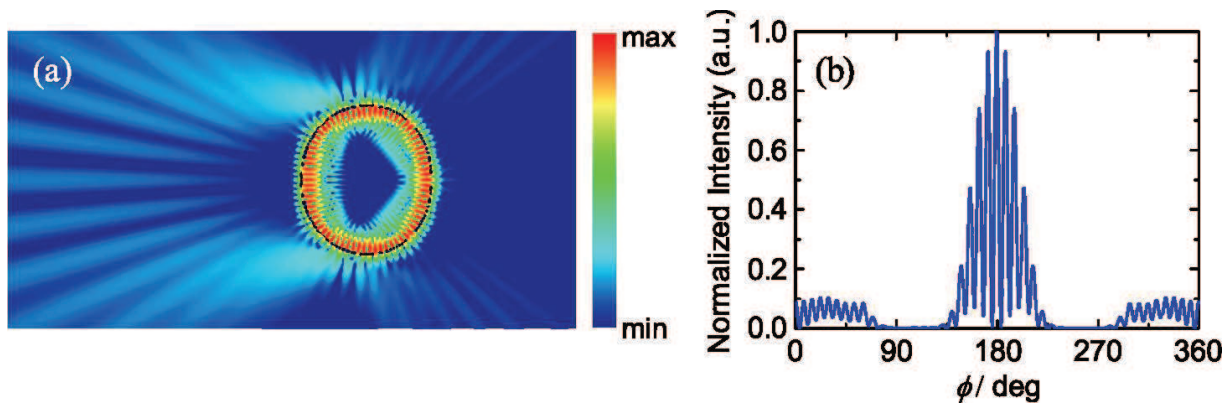}}
\caption{The near field (a) and far-field (b) distributions of $\mathrm{TM}_{37,1}$.
Red-green-blue false colors indicate intensity from high to low in logarithmic scale. }
\end{figure}

Directional emission is essential for free space radiation and excitation
of microcavity, so we should carefully choose a proper cavity. Here,
we focus on the coupling between two cavities, thus the unidirectional
emission is necessary. Different from the most design aiming on a
high ($n=3.3$) \cite{12,14} or a low ($n=1.5$) \cite{13,14} refractive
index, we focus on the cavity with middle refractive index around
$n=2.0$ in which condition the previous cavity design is not applicable,
for the relative refractive index of the semiconductor cavity embedded
in polymer or silica in Ref. \cite{7} and also refractive index of
some materials, such as SiN and diamond, are near 2.0.

Based on the analysis of cavity's symmetry, the cavity shape for high
quality factor and unidirectional emission can be written as \cite{14}
\begin{equation}
B(\phi)=\begin{cases}
R(1-{\textstyle \sum}_{i}a_{i}\cos^{i}\phi), & -\pi/2<\phi\leq\pi/2\\
R(1-{\textstyle \sum}_{i}b_{i}\cos^{i}\phi), & \pi/2<\phi\leq3\pi/2
\end{cases}\label{eq:1}
\end{equation}
where $\phi$ is the polar angle respected to right direction of horizontal
axis, and \textit{R} is the radius . In the frame of ray optics, the boundary $B(\phi)$ of a cavity
with unidirectional emission is optimized by using a hill-climbing
algorithm \cite{16-1}. We obtain a set of parameters for unidirectional
emission. These are $a_{2}=0.03693$, $a_{3}=0.09501$, $b_{2}=0.09791$, $b_{3}=-0.02404$, and zero for other $a_{i}$, $b_{i}$.

Now we turn to wave optics. In the designed ARC, a transverse magnetic
(TM) mode, which has only one no-zero electric field component in
direction perpendicular to the cavity plane, characterized as
$\mathrm{TM}_{37,1}$, which is found through boundary elements method
\cite{15,16}. In the distribution of normalized intensity of this
mode in logarithmic scale {[}Fig. \ref{fig1}(a){]}, two main emission
regions lie in the top and the bottom sites. Through these sites,
the radiation from the mode $\mathrm{TM}_{37,1}$
mainly propagates toward the left direction, and forms a narrow and
high emission peak near the angle of $180^{\circ}$ in the far-field
angular distribution (Fig. \ref{fig1}(b)). From the far-field distribution,
it is found that 54\% of total emission energy is enclosed in the
angle of 40$^{\circ}$ around the main peak or say technically $I_{40}=0.54$
\cite{16-1}.

The dimensionless eigenfrequency of $\mathrm{TM}_{37,1}$
is $kR=22.32621-0.00011i$, where \textit{i} stands for the imaginary
unit. The imaginary part $\mathrm{Im}(kR)$ characterizes the dissipation
of the cavity mode, satisfy the relation
\begin{equation}
\mathrm{Im}(kR)=\frac{-\mathrm{Re}(kR)}{2Q}\label{eq:2}
\end{equation}
 where \textit{Q} is quality factor of a mode. Note that since the
numerical error raised in our calculation (total boundary element
number is 1600) is much larger than $\mathrm{Im}(kR)$, it is hard
to solve the tiny imaginary part $\mathrm{Im}(kR)$ directly {[}6,
17{]}. In this paper, we solve the accurate \textit{Q} with the help
of energy flow method \cite{18}, then obtain the $\mathrm{Im}(kR)$
through Eq. \eqref{eq:2}.

\section{Numerical Results about the Coupling}

\begin{figure}
\label{fig2}\centerline{\includegraphics[clip,width=0.9\linewidth]{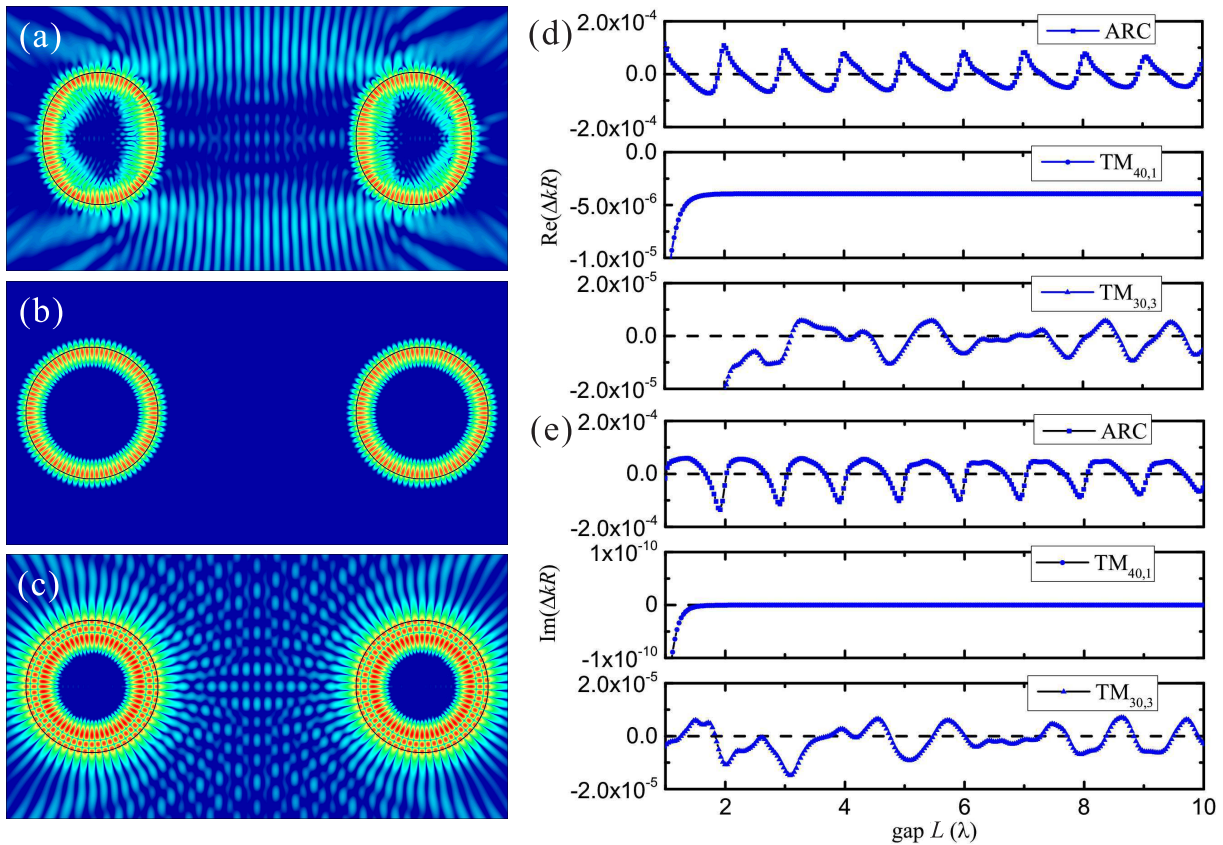}}\caption{Far-field coupling patterns of $\mathrm{TM}_{37,1}$
(a) of the ARC, $\mathrm{TM}_{40,1}$ (b), and
$\mathrm{TM}_{30,3}$ (c) of circular cavity. The
real (d) and imaginary (e) parts of $\Delta kRs$ vary with gap \textit{L}
between two cavities.}
\end{figure}
Place two aforementioned ARCs together with mirror symmetry about
the vertical axis, and let their main emission directions pointing
to each other to form a double-cavities system. Since the system is
linear, we can deduce the dynamics of ARCs through the eigenmodes
of the system. Therefore, we resort eigenvalue equation to investigate
this system rather than research the complicated dynamic process of
coupling directly. When the gap \textit{L} (Fig. \ref{fig3}) between
the cavities is 3\textit{R}, an intensity distribution of the eigenmode,
whose resonant frequency $kR_{\mathrm{EE}}=22.32617-0.00011i$ corresponding
to $\mathrm{TM}_{37,1}$ in single ARC, of the
system is calculated by solving the eigenfunction of $kR$ (Fig. \ref{fig2}(a))
\cite{16}. As we can see in the figure, intensity along the vertical
and horizontal axis of symmetry is maximum, so corresponding complex
amplitude distribution of optical field has reflective symmetry about
the two axes. This mode is classified as even-even (EE) mode. Near
the resonant frequency of EE mode, there always exist other three
classes of mode (odd-even, even-odd, and odd-odd) in the weak coupling
system with four-fold reflective symmetry. In our classification,
the first and last symmetry indicators (E or O) refer to symmetry
state about vertical and horizontal axis, respectively. In Fig. \ref{fig2}(a),
obvious interference fringes with regular periodic oscillation are
shown outside the cavities. Furthermore, the interference fringes
inside the cavities are different to single cavity pattern either
(Fig. \ref{fig1}(a)).

To compare with coupling of circular cavities without directional
emission, two EE modes in coupled circular cavities are calculated
with parameters: gap $L=3R$, radius $R=1$, refractive index $n=2.0$.
The $\mathrm{TM}_{40,1}$ and the $\mathrm{TM}_{30,3}$
modes are found around $kR=22.3$ with $kR_{c1}=22.70596-5.7\times10^{-13}i$
and $kR_{c3}=22.02345-0.00004i$, respectively. Because the \textit{Q}
of $\mathrm{TM}_{40,1}$ mode is as high as $2.0\times10^{13}$,
in other words, little energy can leak out from the cavities. Intensity
outside the cavities is failed to be depicted even in logarithmic
coordinates (Fig. \ref{fig2}(b)). In this case, the coupling between
the modes of two cavities can be ignored, and the field distribution
and the resonant frequency consisted with that of the isolated cavity.
For the $\mathrm{TM}_{30,3}$ mode, its $Q\sim2.5\times10^{5}$
is much less than that of $\mathrm{TM}_{40,1}$
mode, while it is in the same order of magnitude with the $Q\sim1.0\times10^{5}$
of $\mathrm{TM}_{37,1}$ mode of coupled ARCs.
At this point, a regular intensity lattice between the two cavities
is formed by interference. Moreover, the distribution of the regular
pattern is consistent with the interference field between the emissions
of two isolated cavity modes.

Further more, by changing the gap \textit{L} between the two cavities,
we investigate how the resonant frequency of the three EE modes varies
with \textit{L}. In this paper, we only concern the long distance
radiation coupling instead of near distance tunneling coupling, so
the case of $L<\lambda$, which has been well investigated in many
studies of optical molecular, is omitted. Figures \ref{fig2}(d) and
(e) show the real and imaginary part of the \textit{kR} relative to
the isolated one of three modes varying with $L\in\left[\lambda,10\lambda\right]$,
respectively.

The $\mathrm{Re}(\Delta kR)$ and $\mathrm{Im}(\Delta kR)$ of $\mathrm{TM}_{37,1}$
mode of coupled ARCs vary periodically with \textit{L}, and the variation
period is $\lambda$. It suggests that an efficient coupling channel
is established between the ARCs. In addition, the maximum value of
the $\mathrm{Re}(\Delta kR)$ curve shift left about $\lambda/4$
to the $\mathrm{Im}(\Delta kR)$ curve. Their variation amplitudes
are the same in magnitude of $10^{-4}$ at $L=\lambda$, and decrease
slowly as increasing of \textit{L}. Besides, the variation is around
resonant frequency $kR=22.32621-0.00011i$ of isolated cavity, and
the limited frequency at $L=\infty$ tend to the frequency of isolated
cavity also. All above is in line with physical expectations.

Now, we turn to the variation of resonant frequency of coupled circular
cavities. The $\mathrm{Re}(\Delta kR)$ of $\mathrm{TM}_{40,1}$
mode does not show any oscillation in the order of $10^{-6}$, when
the \textit{L} is larger than $1.5\lambda$. As to its $\mathrm{Im}(\Delta kR)$,
the no oscillation order is down to $10^{-11}$. The oscillation of
resonant frequency of $\mathrm{TM}_{30,3}$ mode
is irregular. The amplitude of oscillation is about $10^{-5}$, an
order of magnitude less than that of ARC case.

The oscillation amplitude positively associated with the coupling
strength between the two cavities, even count the ratio of \textsl{Q}s
of$\mathrm{TM}_{30,3}$ to $\mathrm{TM}_{37,1}$,
the coupling strength is enhanced four times in coupled ARCs. We can
draw a conclusion safely that the coupling strength and stability
of ARCs are both better than that of circular cavities because of
the character of the unidirectional emission.

\section{Analysis by the Coupled-Mode Theory}

\begin{figure}
\label{fig3}\centerline{\includegraphics[clip,width=0.5\linewidth]{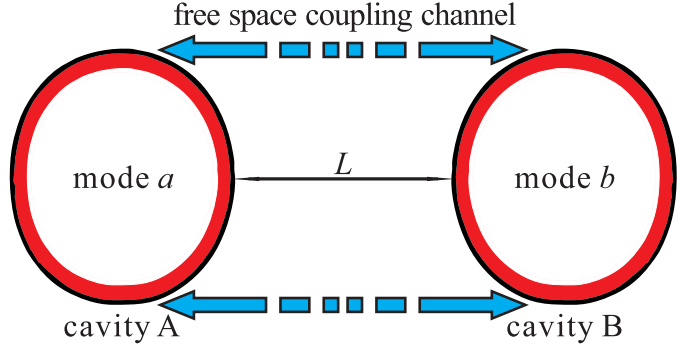}}

\caption{Schematics of coupled cavities.}
\end{figure}

Based on the image of the long distance coupling between ARCs in Fig.
2(a), we can establish a simple model of radiation coupled cavities
{[}see Fig. \ref{fig3}{]}. The eigenmode of separated ARC is described
by the wave function $\Psi_{a(b)}(r)$, with the frequency and intrinsic
loss of $\omega_{a(b)}=\mathrm{Re}(k_{a(b)})\times c$ and $\kappa_{a(b)}=-\mathrm{Im}(k_{a(b)})\times c$,
where \textit{c} is the speed of light in vacuum. For isolated ARCs,
the dynamics of cavity field can be expressed as $\Phi_{a}(r,t)=a(t)\Psi_{a}(r)=a(0)e^{(-i\omega_{a}-\kappa_{a})t}\Psi_{a}(r)$
and $\Phi_{b}(r,t)=b(t)\Psi_{b}(r)=b(0)e^{(-i\omega_{b}-\kappa_{b})t}\Psi_{b}(r)$,
with dimensionless coefficients of $a(t)$ and $b(t)$ .

With weak coupling perturbation, the wave function of a composite
system can be written as $\Phi(r,t)=a(t)\Psi_{a}(r)+b(t)\Psi_{b}(r)$
\cite{19}. Considering the radiation coupling between two ARCs, and
according to the coupled-mode theory, the dynamics of the system satisfy
the following differential equations
\begin{equation}
\begin{cases}
\frac{d}{dt}a(t)=(-i\omega_{a}-\kappa_{a})a(t)+ge^{i\theta}b(t)\\
\frac{d}{dt}b(t)=(-i\omega_{b}-\kappa_{b})b(t)+ge^{i\theta}a(t)
\end{cases}\label{eq:3}
\end{equation}
where $ge^{i\theta}$ is the term of radiation coupling , in which
\textit{$g\geq0$} is the coupling strength and \textit{$\theta$}
is the phase shift that the coupling wave undergoes from a cavity
to another.

According to the coupled mode equations, we can solve the new normal
modes of the composite system as $\Psi_{\pm}(r)=c_{a}\Psi_{a}(r)+c_{b}\Psi_{b}(r)$,
which frequency $\omega_{\pm}$ and coefficients $\left(\begin{array}{c}
c_{a}\\
c_{b}
\end{array}\right)$ are just the eigenvalues and eigenvectors of the matrix
\begin{equation}
M=\left(\begin{array}{cc}
\omega_{1} & \gamma\\
\gamma & \omega_{2}
\end{array}\right)=\left(\begin{array}{cc}
-i\omega_{a}-\kappa_{a} & ge^{i\theta}\\
ge^{i\theta} & -i\omega_{b}-\kappa_{b}
\end{array}\right).
\end{equation}
 It is easy to solve the eigenfrequency
\begin{equation}
\omega_{\pm}=\frac{1}{2}(\omega_{1}+\omega_{2}\pm\sqrt{(\omega_{1}-\omega_{2})^{2}+4\gamma^{2}}),
\end{equation}
and corresponding coefficients
\begin{equation}
\left(\begin{array}{c}
c_{a}\\
c_{b}
\end{array}\right)=\frac{1}{\sqrt{|\omega_{\pm}-\omega_{1}|^{2}+|\gamma|^{2}}}\left(\begin{array}{c}
\gamma\\
\omega_{\pm}-\omega_{1}
\end{array}\right).\label{eq:6}
\end{equation}

\begin{figure}
\label{fig4}\centerline{\includegraphics[clip,width=1\linewidth]{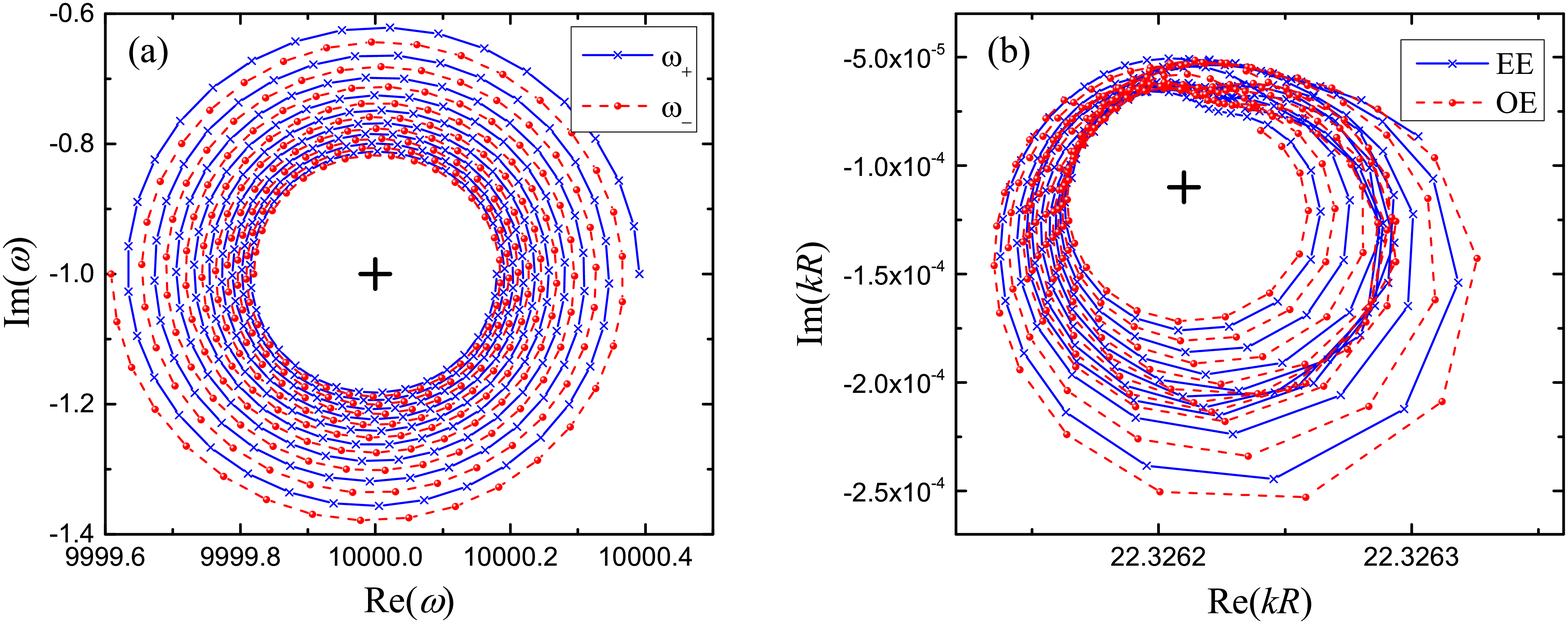}}

\caption{The tracks of \textit{kR} theory (a) and simulation (b). Center cross
indicate the location of resonant frequency without coupling.}
\end{figure}

For example, in the case of two identical ARCs, i.e. $\omega_{a}=\omega_{b}$
and $\kappa_{a}=\kappa_{b}$, we have $\omega_{\pm}=-i\omega_{a}-\kappa_{a}\pm ge^{i\theta}$
and $\Psi_{\pm}=\frac{1}{\sqrt{2}}(\Psi_{a}\pm\Psi_{b})$. To the
radiation coupling, the phase \textit{$\theta$} is proportional to
the propagation distance \textit{L} as $\theta=\frac{L}{\lambda}\times2\pi+\theta_{0}$,
where $\theta_{0}$ is constant phase according to coupling phase
when $L=0$. Therefore, when the \textit{$\theta$} increases with
the increasing of \textit{L}, the points of $\omega_{\pm}$ draw a
circle centered at $\omega_{1}$ and with radius \textit{g} in the
complex plan of \textit{$\omega$}. Noting that a pair of $\omega_{+}$
and $\omega_{-}$ always locate at the ends of a diameter of the circle.
Additionally, \textit{g} is not a constant but a function of \textit{L}.
Since \textit{g} stands for the magnitude of coupling, i.e. a portion
of emitting energy from one cavity coupled into the other cavity,
we can assume the function as
\begin{equation}
g(L)=g(0)/\sqrt{L+2R}.\label{eq:7}
\end{equation}
The magnitude of \textit{g} must less than that of $\mathrm{Im}(\omega_{1})$,
which refers to the emission amount of one cavity. Consequently, the
data range of \textit{g} is $[0,1]\times\mathrm{Im}(\omega_{1})$.
Let us give an example. Assuming $\theta_{0}=0$, $\lambda=0.3$,
$g(0)=0.9$, and $L\in[1,10]\times\lambda$, we get the double spiral
curves of $\omega_{\pm}$ in \textit{$\omega$} plane (Fig. \ref{fig4}(a)),
which starts from the outer points and gradually approaches the center
cross $\omega_{1}$ with increasing of \textit{L}.

The $\omega_{+}$ is associated with \textit{k,} which is calculated
in the last section, with proportional relationship $\omega_{+}=-ikc$.
Transporting the curves of ARCs in Fig. \ref{fig2} to a spiral curve
in \textit{kR} complex plane, we get the \textit{kR} track of varying
with \textit{L} (Fig. \ref{fig4}(b) blue solid line). The point on
the spiral curve closes to center cross with increasing \textit{L},
which indicates that the coupling strength is not a constant but a
decreasing value too. In addition, the density of points in upper
part of the curve is higher than that in the under part. It reveals
that there is no strictly linear relationship between the $\theta$
and the \textit{L}. Furthermore, a \textit{kR} track of odd-even mode
corresponding to $\omega_{-}$ is calculated (Fig. \ref{fig4}(b)
red dash line). Again, it is a spiral curve like the track of EE mode,
and intertwined with the spiral line of EE mode around the center
cross.

Though the pictures of analytical and numerical results are qualitatively
consistent with each other, the last one is more complicated. The
reason is the real coupling term $ge^{i\theta}$ as a function of
\textit{L} is more complex than we assumed. For example, not like
the waveguide coupling case, in free space coupling case, coupling
phase \textit{$\theta$} is related with the shape and location of
the beam emitted by isolated cavity. So, the linear relationship with
\textit{L} is a simplified model.

Before a further discussion, we evaluate the $g$ from the data shown
in Fig. \ref{fig4}(b), for $2g=|\omega_{+}-\omega_{-}|$. As is shown
in Fig. 5(a) (solid blue line), when the \textit{L} increases the
envelope of $g$ decreases slowly. Obviously, a relatively high $g$
($>0.4$) last in a wide range to $L=10\lambda$. Besides, a fitted
line (Fig. 5(a) dash red line) according to Eq. \eqref{eq:7} agrees
with simulation data in certain degree, if we ignore the oscillation.
Here, the oscillation stems from multi-modes combination induced by
multi-channel coupling in free space \cite{11}, which is not included
in our simplified model of single mode coupling.

\section{Dynamics of Coupled ARCs}

\subsection{On-resonance}

Since the eigenfunction $\Phi_{\pm}$ of coupled mode is known, the
time evolution of any complex function $\Phi$ in the subspace can
be determined. Here, as an example, we set the initial state of coupled
system to $\Phi_{a}$, and discuss the energy transportation between
the two coupled cavities.

When $t=0$, we have $\Phi(r,0)=\Phi_{a}(r,0)=[\Phi_{+}(r,0)+\Phi_{-}(r,0)]/\sqrt{2}$.
Later the state function at time \textit{t} reads
\begin{align}
\Phi(r,t) & =\frac{1}{\sqrt{2}}[\Phi_{+}(r,0)e^{\omega_{+}t}+\Phi_{-}(r,0)e^{\omega_{-}t}]
\end{align}
To find the energy in the two cavities, last formula can be rewritten
as:
\begin{align}
\Phi(r,t) & =\frac{1}{2}[(e^{\omega_{+}t}+e^{\omega_{-}t})\Psi_{a}+(e^{\omega_{+}t}-e^{\omega_{-}t})\Psi_{b}]
\end{align}
The normalized energy in cavity A or B reads
\begin{align}
E_{a,b} & =\left|\frac{1}{2}(e^{\omega_{+}t}\pm e^{\omega_{-}t})\right|^{2}\nonumber \\
 & =\frac{1}{2}e^{-2\kappa_{a}t}[\cosh(2g\cos\theta\cdot t)\pm\cos(2g\sin\theta\cdot t)]\label{eq:4}
\end{align}
where positive and negative sign correspond to a and b, respectively.

\begin{figure}
\label{fig5}\centerline{\includegraphics[clip,width=0.9\linewidth]{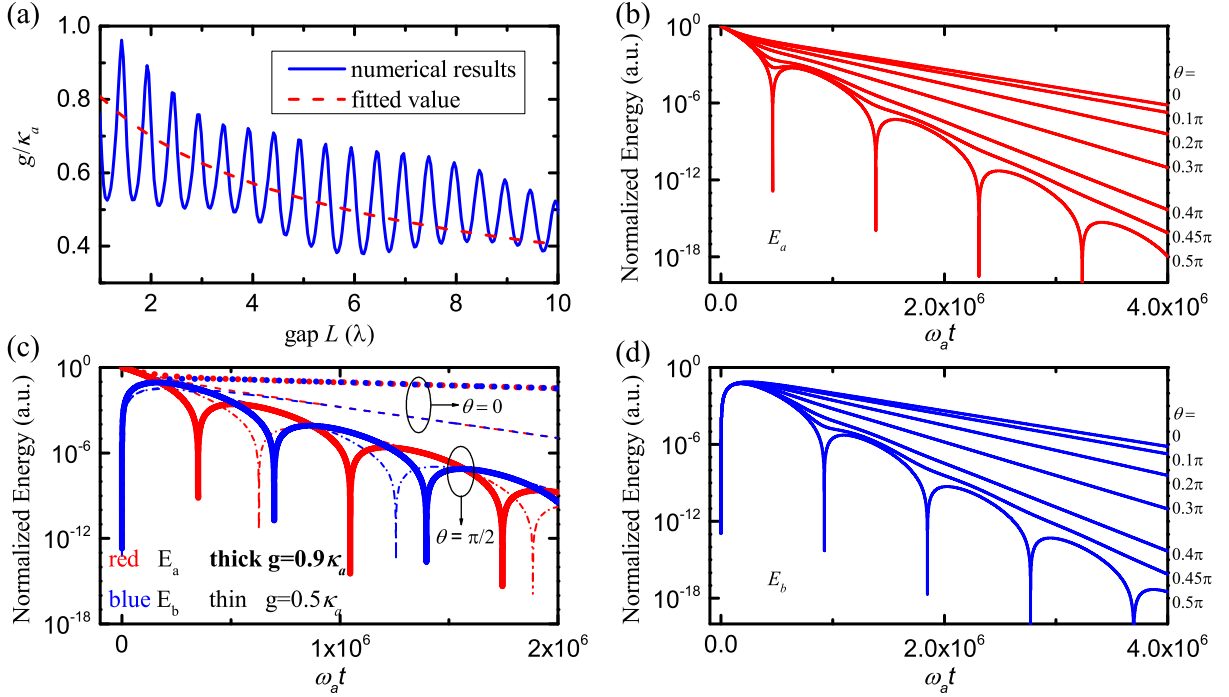}}

\caption{(a) The values of $g/\kappa_{a}$ get from simulation results (solid
blue line) and Eq. \eqref{eq:7} with $g(0)=1.4$ (dash red line).
Time evolution of normalized energy of two coupled cavities. Energy
in cavity A (b) and B (d) vary with time when $g=0.68\kappa_{a}$
and $\theta=0,\,0.1\pi,\,0.2\pi,\,0.3\pi,\,0.4\pi,\,0.45\pi,\,0.5\pi$.
(c) Evolution of energy at different combination of parameters.}

\end{figure}

Taking some examples, the mode $\mathrm{TM}_{37,1}$
in isolated ARC has a $Q\sim10^{5}$, i.e. $\kappa_{a}=5\times10^{-6}\omega_{a}$.
We pick a middle $g=0.68\kappa_{a}$ and observe the time evolution
of energy in cavity A and B at some coupling angles of $\theta$(Figs.
5(b) and (d)). When $\theta=0$, the curves show there is already
considerable energy transported from cavity A into cavity B at the
time $\omega_{a}t\sim2.5\times10^{5}$, though the initial energy
is wholly in cavity A. Then after a period of time the energy is balanced
in both cavity, and attenuates exponentially with slope of $-2(\kappa_{a}-g)/\omega_{a}$.
Enlarge the $\theta$ gradually, the feature of transportation of
energy is similar with case of $\theta=0$ except the final attenuation
slope which changes to $-2(\kappa_{a}-g\cos\theta)/\omega_{a}$. When
the $\theta$ close to $\pi/2$, such as $0.45\pi$, the augment $2g\cos\theta\cdot t$
of attenuation (first) term bracketed in Eq. \eqref{eq:4} is small,
while the oscillating (second) term plays a major role when t is not
too large. At this time, energy oscillates between the two cavities
several times, then the amplitude of the oscillation reduces to zero,
and energy becomes equal in two cavities. When the $\theta$ equals
$\pi/2$, only the oscillating term left in square bracket of Eq.
\eqref{eq:4}. Thus, energy attenuates with oscillation (Fig. 5 solid
line), that is to say the energy in the two cavities empty in turn.
The attenuation slope reach the maxima value $-2\kappa_{a}/\omega_{a}$.
The case of $\theta\in[\pi/2,\pi]$ repeats the result in interval
$[0,\pi/2]$ with the reversal manner $\pi-\theta$. Furthermore,
$\pi$ is the period of the dynamic rule of the evolution of modes
varying with $\theta$.

Then we change the $g$ to $0.5\kappa_{a}$ and $0.9\kappa_{a}$ for
comparison (Fig. 5(c)). To the smaller $g=0.5\kappa_{a}$ case, the
speed of energy exchange slows down, and the time requirement for
energy balance in coupled cavities increases. Moreover, the proportion
of the energy transported from one cavity to another decreases, i.e.
more energy emits out of the coupling system. Hence, the slope of
energy attenuation line decreases except the constant slope in $\theta=\pi/2$
case. If $\theta=\pi/2$, the period of energy emptying in cavities
increases with decreasing of \textit{g}. On the contrary, if we change
the mode used in coupling with different \textit{Q}, then the time
to achieve energy balance is proportional to the \textit{Q}.

\subsection{Off-resonance}

\begin{figure}
\label{fig6}\centerline{\includegraphics[clip,width=0.9\linewidth]{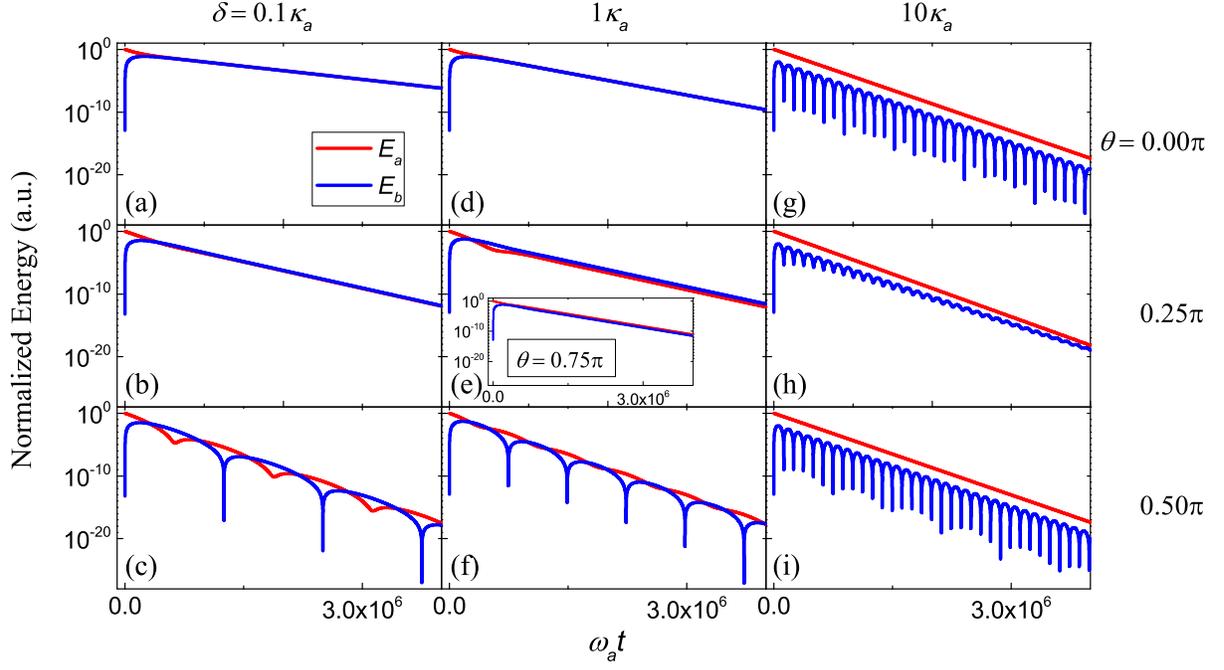}}

\caption{Time evolution of normalized energy of two mismatched cavities. Three
columns, from left to right, week, middle, and strong represent detuning,
respectively. Each row cope with a typical value of $\theta$. Insert
in (c): middle detuning and with $\theta=0.75\pi$.}
\end{figure}

In practice, two coupled cavities may not be identical but have slight
different geometry, so the resonant frequencies of cavities have a
small mismatch, i.e. $\omega_{1}\neq\omega_{2}$. Keeping $\kappa_{a}$
and $\kappa_{b}$ equal, we study the detuning $\delta=\omega_{a}-\omega_{b}$
influence on the dynamics of the coupling. Then, like in the on-resonance
case, after derivation from Eq. \eqref{eq:6} we get the evolution
equation of energy and depict it in Fig. 6 with $g=0.68\kappa_{a}$
and $Q=10^{5}$ again.

If the detuning induced by size difference is slight, say $\delta=0.1\kappa_{a}$,
the resonances in two cavities overlap significantly with each other.
Then, the resonant frequency of coupled mode still varies around the
average value of isolated frequencies. Yet the reduction of the coupling
strength shrinks the splitting of two coupled modes. Besides, energy
of modes distribute unequally in two cavities after a transient state.
When $\theta$ increases from 0 to $0.5\pi$, the unbalance of the
energy distribution becomes worse and the slope of the decay curve
becomes steeper (Fig. 6 first and second columns). For the Fig. 6(c),
though the cavity A keep more energy than cavity B in the oscillation,
the oscillation period is nearly as same as that in on-resonance case
(Fig. 5(c) thin line).

The imbalance of energy distribution becomes more serious when the
detuning $\delta$ gets larger. In the middle detuning case $\delta=1\kappa_{a}$
(Fig. 6 second column), we found whether the over coupling or under
coupling occurs depends on the coupling phase $\theta$. For instance,
when $\theta=0.25\pi$ the energy flow from cavity A to cavity B can
not reflow back to cavity A completely, but when $\theta=0.75\pi$
the energy in cavity A can not transport to B entirely (Fig. 6 middle
panel and its insert). In addition, the evolution of energy about
the $\theta$ is also in the period of $\pi$ but is different in
the cases of $\theta$ with of $\pi-\theta$ .

In fact, when the detuning is much larger than the linewidth, say
$\delta\gg\kappa_{a}$, the resonated mode in one cavity is merely
influenced by another cavity (Fig. 6 last column). In other words,
energy cannot deliver from cavity A to cavity B efficiently, so mode
\textit{a} decays exponentially without oscillation. At this time,
a very few energy in cavity B oscillates intensely, as there arises
a beat oscillation, whose period is inversely proportional to $\delta$.

\section{Conclusion}

In summary, an efficient coupling in long distance is established
by employing unidirectional ARCs. We designed a unidirectional cavity
with refractive index 2.0 for coupling. Comparing with the free-space
far-field coupling of two circular cavities, the coupling of ARCs
is strong and stable, which makes ARC being the first choice for far-field
coupling. The coupled mode theory works well on solving the coupled
problem based on the isolated mode. Results show that the coupled
mode is the superposition of two single modes, and the corresponding
resonant frequency is the sum (difference) of single mode frequency
and a coupling term. Moreover, based one the result of the coupled
mode theory, the time revolution of the light field, which is described
by an arbitrary complex function, can be calculated by decomposing
it with bases of normal mode. Then under on- and off-resonance coupling
conditions, the transmission of energy in the coupling system is depicted
in time domain. It need to note that coupling mode or time domain
characters both extremely rely on the coupling phase $\theta$. So
far we have solved the low coupling efficiency problem encountered
in Ref. \cite{7}.This paper only discusses two coupled passive ARCs.
Next, coupling system composed of a laser cavity and a passive cavity
\cite{8} in ARC form need to be studied, and the coupling of multi-cavities
\cite{key-2} is interesting too.

\section*{Acknowledgments}
This work was supported in part by the National Natural Science Foundation
of China under Grant 11204169 and 11247289. CLZ and FWS were supported
by the 973 Programs (No. 2011CB921200), the National Natural Science
Foundation of China (NSFC) (No. 11004184), the Knowledge Innovation
Project of the Chinese Academy of Sciences (CAS).



\end{document}